\documentclass[11pt]{article}
\usepackage[colorlinks=true,citecolor=blue,linkcolor=blue]{hyperref}
\usepackage[left=2cm,right=2cm,top=2cm,bottom=2cm]{geometry}
\usepackage[utf8]{inputenc}
\usepackage{cite}
\usepackage{psfrag}
\usepackage{graphicx}
\usepackage{graphics}
\usepackage{epsfig}
\usepackage{amsmath,amsfonts,amssymb}
\usepackage{euscript}
\usepackage{pstricks}
\usepackage{wrapfig}
\usepackage{slashed}
\usepackage[toc,page]{appendix}
\usepackage{here}
\usepackage{hyperref}
\usepackage{booktabs}
\usepackage{multirow}
\usepackage{siunitx}
\hypersetup{
    colorlinks=true,
    linkcolor=blue,
    filecolor=magenta,      
    urlcolor=cyan,
    }

\date{}

\begin{document}
	\title{\vspace{-3cm}
		\hfill\parbox{4cm}{\normalsize \emph{}}\\
		\vspace{1cm}
		{Effect of electron spin polarization in laser-assisted electron-proton scattering}}
	\vspace{2cm}

\author{I. Dahiri$^{1}$, M. Baouahi$^{1}$, M. Jakha$^1$, S. Mouslih$^{1,2}$ B. Manaut$^{1}$, S. Taj$^{1,}$\thanks{Corresponding author, E-mail: s.taj@usms.ma}  \\
		{\it {\small$^1$ Polydisciplinary Faculty, Laboratory of Research in Physics and Engineering Sciences,}}\\ {\it{\small Team of Modern and Applied Physics, Sultan Moulay Slimane University, Beni Mellal, 23000, Morocco.}}\\	
		{\it {\small$^2$Faculty of Sciences and Techniques,
		Laboratory of Materials Physics (LMP),
		Beni Mellal, 23000, Morocco.}}
		}
	\maketitle \setcounter{page}{1}
\date{\today}
\begin{abstract}
This work aims to study theoretically the electron-proton scattering for initially spin-polarized electrons in the presence of a circularly polarized electromagnetic field. Using the first Born approximation and the Dirac-Volkov states for dressed electrons, we derive an analytic expression for the differential cross-section with the help of the spin-helicity formalism. Meanwhile, a well-known concept of spin-flip and spin non-flip differential cross sections is applied. The influence of the
electromagnetic field on the degree of polarization of the scattered electron is investigated, and the effect of electron spin polarization is examined. We found that the initial spin of the electron can be affected depending on its kinetic energy during the scattering process. This work is a continuation of a recent paper in which we studied the unpolarized electron-proton scattering in the presence of a laser field (Dahiri, et al., Laser Phys. Lett. 18 (2021) 096001). 
\end{abstract}
Keywords: spin effects, QED calculation, laser-assisted processes
\section{Introduction}\label{Sec.0}
In view of the new developments in laser technology, high intensity and short duration laser pulses are generated nowadays in the laboratory \cite{Laser_technologies,Laser_technologies_1,Laser_technologies_2}.  This has greatly inspired researchers to study various ultrafast processes in the presence of strong laser fields, such as atomic processes \cite{ioniz2s,taj2019,idrissi2013,taj2011,manaut2009,Manaut_2005,Attaourti} and those that occur in the frameworks of quantum electrodynamics (QED) \cite{Dahiri_2021,Mekaoui2021,Narozhny_2015,DiPiazza_2012,Ehlotzky_2009} and electroweak theory \cite{oualichin,ElAsri2021,pidecay,wdecay,kdecay,zdecay}, as well as recently the processes of Higgs boson production beyond the Standard Model \cite{Ouali2021,ouhammouchin,Ouhammou2021}. New experimental perspectives for the study of these processes in the presence of laser field are proposed by future high field laboratories, such as the Extreme Light Infrastructure (ELI) \cite{Mourou_2011} and the Exawatt Center for Extreme Light Studies (XCELS) \cite{XCELS}. The majority of theoretical studies investigating scattering processes in QED treat the involved particles as unpolarized and therefore their spins are averaged over. At high energies, the scattering of spin-polarized particles is an important aspect of physics \cite{schwartz}, owing much to the development of new detectors and sources of such particles. Polarized beams and targets are used to study various effects of spin physics. As a definition, an ensemble of particles (e.g., electrons) is said to be polarized if the electron spin has a preferential orientation so that there exists a direction for which the two possible spin states are not equally populated. In early experiments with free electrons, the direction of their spins was rarely considered. Whenever the spin direction plays a role, one has to average over all spin polarizations in order to describe the experiments properly. It has long been possible to produce electron beams in which the spin has a preferential orientation \cite{clendenin,hodge,pierce1,pierce2}. There are many reasons for the interest in polarized electrons. The most important one is that in physical experiments one seeks to define as accurately as possible the initial and/or final states of the considered systems \cite{kessler}. Several theoretical studies have considered the effects of spin polarization in various processes. Deep inelastic scattering of polarized leptons and protons revealed interesting details on the  spin-structure of the proton's  constituents \cite{anselmino95,steven2009}, and the parity  non-conservation effects have been investigated by using polarized beams  \cite{prescott78,labzowsky2001}. The polarization effects on Compton scattering have been studied by several works \cite{compton1,compton2,compton3}. Mott scattering of polarized electrons has been studied in a strong laser field of circular \cite{manaut2009} and linear \cite{Manaut_2005} polarization.  The unpolarized scattering of electrons by protons is one of the basic processes in QED that has been widely studied, as it is considered the simpler way to access more detailed information on the internal structure of proton. The effect of a circularly polarized laser field on this process has been the focus of a recent paper for unpolarized electrons \cite{Dahiri_2021}. As an extension of this paper, the present work aims to study the effect of electron spin polarization on the electron-proton scattering process in the presence of a laser field using the concept of a spin-polarized relativistic Dirac particle. Polarization effects in elastic proton-electron scattering have been investigated in \cite{gakh} using polarized beams for both electron and proton. The same approach adopted here has been used to study laser-assisted processes in relativistic atomic physics \cite{manaut2009,Manaut_2005,idrissi2013}. The organization of this paper is as follows: in section \ref{Sec.1}, we develop a theoretical analytical calculation of the differential cross section (DCS) of electron-proton (e-p) scattering in the absence and in the presence of the laser field for both unpolarized and polarized electrons. Then, in section \ref{Sec2}, we will discuss the different results obtained and we end with a conclusion in section \ref{Sec3}. We mention that this work is developed in the laboratory system of proton and for a space-time metric of signature $ (+,-,-,-) $. We use the atomic unit system $\hbar=e=m_{e}=1$ during the theoretical calculation.
\section{Overview of the theory}\label{Sec.1}
In the theoretical treatment, we will present an analytical calculation of the free DCS in the first Born approximation. Then, we introduce the concept of spin polarization, and at the end the expression of the DCS in the presence of a monochromatic laser field is derived based on the approach of describing the electrons by the Volkov states \cite{volkov} and the spin projection formalism.
\subsection{Laser-free unpolarized differential cross section}
\begin{figure}[ht]
\centering
    \includegraphics[scale=0.3]{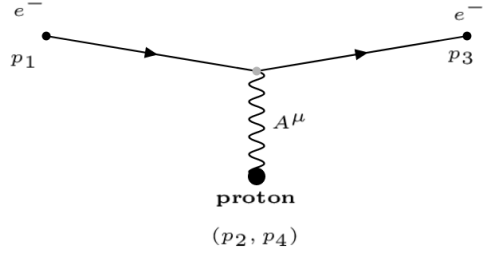}\par
\caption{Electron-proton scattering in the laboratory system, with $A^{\mu}$ is the four-potential produced by the proton.} \label{Figure:0}
\end{figure}

The elastic scattering of a proton by electron impact, shown in Fig.~\ref{Figure:0}, is studied by considering the proton as a structureless Dirac particle.
In Fig.~\ref{Figure:0}, the $p_{n}$ (for $n=\{1,3\}$) represent the 4-momentum of the incident ($p_{1}$) and scattered ($p_{3}$) electrons, while $p_{m}$ (for $m=\{2,4\}$) represent the 4-momentum of the proton in its initial ($p_{2}$) and final ($p_{4}$) states. In the absence of the laser field, the particles studied are described by Dirac wave functions, normalized by the quantum volume $V$, as follows:
\begin{eqnarray}\label{Dirac_libre}
\left\lbrace\begin{array}{c}
\psi_{p_{n}}(x)=\frac{1}{\sqrt{2E_{n}V}}u(p_{n},s_{n})e^{-ip_{n}.x},\\
\, \\
~~\psi_{p_{m}}(y)=\frac{1}{\sqrt{2E_{m}V}}u(p_{m},s_{m})e^{-ip_{m}.y},
\end{array}\right.
\end{eqnarray}
$E_{n,m}$ and $s_{n,m}$ represent the energy and the spin of the particles respectively. At the leading order, the matrix element $S_{fi}$ representing our scattering process is:
\begin{equation}\label{Smatrix_libre_0}
S_{fi}=-i\int d^{4}x~\overline{\psi}_{p_{3}}(x)\slashed{A}(x)\psi_{p_{1}}(x),
\end{equation}
with $\slashed{A}(x)=A_{\mu}(x)\gamma^{\mu}$, where $A_{\mu} $ is the electromagnetic 4-potential of the scattering given by:
\begin{eqnarray}\label{4-potentiel}
A_{\mu}(x)&=&-\int d^{4}y~D_{F}(x-y)\bar{\psi}_{p_{4}}(y)\gamma_{\mu}\psi_{p_{2}}(y), \nonumber \\
&=& \int d^{4}y~\frac{d^{4}q}{(2\pi)^{4}}\frac{e^{-iq(x-y)}}{q^{2}+i\epsilon}\big[\bar{\psi}_{p_{4}}(y)\gamma_{\mu}\psi_{p_{2}}(y)\big],
\end{eqnarray}
such as the element $D_{F}(x-y)=-\int d^{4}qe^{-iq(x-y)}/\big[(2\pi)^{4}(q^{2}+i\epsilon)\big]$ is the Feynman electromagnetic propagator. By replacing Eqs.~(\ref{Dirac_libre}) and (\ref{4-potentiel}) in the expression of $S_{fi}$ (\ref{Smatrix_libre_0}), the matrix element becomes:
\begin{eqnarray}\label{Smatrix1}
S_{fi}&=&\frac{-i}{\sqrt{16E_{1}E_{2}E_{3}E_{4}V^{4}}}\int d^{4}x~d^{4}y~\frac{d^{4}q}{(2\pi)^{4}}\frac{e^{i(p_{3}-p_{1}-q).x}e^{i(p_{4}-p_{2}+q).y}}{q^{2}+i\epsilon}\mathcal{M}_{fi}, \nonumber\\
&=&\frac{-i}{\sqrt{16E_{1}E_{2}E_{3}E_{4}V^{4}}}\frac{(2\pi)^{4}\delta^{4}(p_{4}-p_{2}+p_{3}-p_{1})}{(p_{3}-p_{1})^{2}+i\epsilon}\mathcal{M}_{fi},
\end{eqnarray}
where the element $\mathcal{M}_{fi}$ represents the scattering amplitude such that:
\begin{eqnarray}
\mathcal{M}_{fi}=\big[\bar{u}(p_{3},s_{3})\gamma^{\mu}u(p_{1},s_{1})\big]\big[\bar{u}(p_{4},s_{4})\gamma_{\mu}u(p_{2},s_{2})\big].
\end{eqnarray}
To express the unpolarized DCS, the squared matrix element $|S_{fi}|^2$ is weighted by the flux of the incident electrons, the phase space of the particles in the final state and by unit of time $T$. Thus, we get:
\begin{eqnarray}\label{DCS_0}
d\bar{\sigma}=\frac{Vd^{3}\vec{p}_{3}}{(2\pi)^{3}}\frac{Vd^{3}\vec{p}_{4}}{(2\pi)^{3}}\frac{1}{|J_{\text{inc.}}|T}\frac{(2\pi)^{4}VT\delta^{4}(p_{4}-p_{2}+p_{3}-p_{1})}{16E_{1}E_{2}E_{3}E_{4}V^{4}q^{4}}|\overline{\mathcal{M}_{fi}}|^{2},
\end{eqnarray}
where we average over the initial spins and sum over the final ones as follows:
\begin{equation}\label{sum-moy}
|\overline{\mathcal{M}_{fi}}|^{2}=\frac{1}{4}\sum_{s_{n,m}}|\mathcal{M}_{fi}|^{2}.
\end{equation}
The incident particle flux is $|J_{\text{inc.}}|=|\vec{p}_{1}|c^{2}/(E_{1}V)$. To integrate with respect to $d^{3}\vec{p}_{4}$, we will use the following formula: 
\begin{eqnarray}
\frac{d^{3}\vec{p}_{4}}{E_{4}}=2\int_{-\infty}^{+\infty}d^{4}p_{4}\delta(p_{4}^{2}-M^{2}c^{2})\Theta(p_{4}^{0});\quad\,\,\Theta(p_{4}^{0})=\left\lbrace\begin{array}{ccc}
&1&~~\text{for}~~p_{4}^{0}>0\\
&0&~~\text{for}~~p_{4}^{0}<0,
\end{array}\right. 
\end{eqnarray}
where $M$ is the rest mass of the proton.
Using $d^{3}\vec{p}_{3}=|\vec{p}_{3}|^{2}d|\vec{p}_{3}|d\Omega_{f}$ and $E_{3}dE_{3}=c^{2}|\vec{p}_{3}|d|\vec{p}_{3}|$, we find that the unpolarized DCS is expressed as : 
\begin{eqnarray}\label{DCS_1}
\frac{d\overline{\sigma}}{d\Omega_{f}}=\frac{1}{8(2\pi)^{2}c^{6}Mq^{4}}\frac{|\vec{p}_{3}|}{|\vec{p}_{1}|}\int \delta(p_{4}^{2}-M^{2}c^{2})dE_{3}|\overline{\mathcal{M}_{fi}}|^{2}.
\end{eqnarray}
The integration over $dE_{3}$ is performed using the following property \cite{greiner}: 
\begin{eqnarray}\label{property_1}
\int dx~f(x)\delta(g(x))=\frac{f(x)}{|g'(x)|}\bigg|_{g(x)=0}.
\end{eqnarray}
In our case: $g'(E_{3})=M+\frac{E_{1}}{c^{2}}-\frac{E_{3}}{c^{2}}\frac{|\vec{p}_{1}|}{|\vec{p}_{3}|}F(\theta_{i},\theta_{f},\varphi_{i},\varphi_{f})$, where
\begin{eqnarray}
F(\theta_{i},\theta_{f},\varphi_{i},\varphi_{f})=\sin(\theta_{i})\sin(\theta_{f})\big[\cos(\varphi_{i})\cos(\varphi_{f})+\sin(\varphi_{i})\sin(\varphi_{f})\big]+\cos(\theta_{i})\cos(\theta_{f}).
\end{eqnarray}
Finally, we give the expression of the unpolarized DCS as follows:
\begin{eqnarray}\label{DCS_free}
\bigg(\frac{d\overline{\sigma}}{d\Omega_{f}}\bigg)^{\text{laser-free}}_{\text{unpolarized}}=\frac{1}{16(2\pi)^{2}Mc^{6}q^{4}}\frac{|\vec{p}_{3}|}{|\vec{p}_{1}|}\frac{|\overline{\mathcal{M}_{fi}}|^{2}}{|g'(E_{3})|},
\end{eqnarray}
where
\begin{eqnarray}\label{trace1}
|\overline{\mathcal{M}_{fi}}|^{2}=\frac{1}{4}\text{Tr}\big[(c\slashed{p}_{3}+c^{2})\gamma^{\mu}(c\slashed{p}_{1}+c^{2})\gamma^{\nu}\big]\text{Tr}\big[(c\slashed{p}_{4}+Mc^{2})\gamma_{\mu}(c\slashed{p}_{2}+Mc^{2})\gamma_{\nu}\big].
\end{eqnarray}
\subsection{Laser-free differential cross section for polarized electrons}
Until date, all calculations have been done assuming that the electron's spin has not been detected. Now, we introduce the calculation of the DCS for polarised electrons. As a first step, we define the spin 4-vector $S_{\lambda}^{\mu}$ as:
\begin{eqnarray}
S_{\lambda}^{\mu}=\frac{\lambda}{c}\Big(|\vec{p}|,\frac{E}{c}\hat{p}\Big),~~~~\text{where}~~~\,\,\hat{p}=\frac{\vec{p}}{|\vec{p}|},
\end{eqnarray}
which verifies the normalization relation $S^{\mu}S_{\mu}=-1$, and the orthogonality relation with the 4-momentum $p$ such that: $(p.S)=p^{\mu}.S_{\mu}=0$. The parameter $\lambda$ defines the spin orientation (spin Up for $\lambda=+1$ and Down for $\lambda=-1$).\\
Subsequently, the spin projection operator is introduced:  
\begin{eqnarray}
\hat{\Sigma}(S)=\frac{1}{2}(1+\gamma_{5}\slashed{S}),~~~\quad\text{with}~~~ \gamma_{5}=i\gamma^{0}\gamma^{1}\gamma^{2}\gamma^{3}=-i\gamma_{0}\gamma_{1}\gamma_{2}\gamma_{3}.
\end{eqnarray} 
The spin projection operator $\hat{\Sigma}(S)$ verifies the following property:
\begin{eqnarray}
\hat{\Sigma}(S)u(p,S')=\delta_{SS'}u(p,S').
\end{eqnarray}
If we consider the spin polarization of electrons, the only quantity that changes in the expression of the laser-free unpolarized DCS (\ref{DCS_free}) is $|\overline{\mathcal{M}_{fi}}|^{2}$ because it is the only part that depends on the spin. It becomes
\begin{equation}\label{sum-moy1}
|\overline{\mathcal{M}_{fi}}|^{2}=\frac{1}{2}\sum_{s_{2},s_{4}}|\mathcal{M}_{fi}|^{2}.
\end{equation}
In Eq.~(\ref{sum-moy1}), one has to be careful since there are no
summations on electron spin (either initial or final) polarizations. The factor $\tfrac{1}{2}$ is due to the average over the initial polarizations of the unpolarized proton. Introducing the two spin projection operators for the initial and final electrons, we find
\begin{equation}\label{trace2}
\begin{split}
|\overline{\mathcal{M}_{fi}}|^{2}=&\text{Tr}\Big[\big(c\slashed{p}_{3}+c^{2}\big)\gamma^{\mu}\Big(\frac{1+\lambda_i\gamma_{5}\slashed{S}_{i}}{2}\Big)\big(c\slashed{p}_{1}+c^{2}\big)\gamma^{\nu}\Big(\frac{1+\lambda_f\gamma_{5}\slashed{S}_{f}}{2}\Big)\Big]\\
&\times \dfrac{1}{2} \text{Tr}\big[(c\slashed{p}_{4}+Mc^{2})\gamma_{\mu}(c\slashed{p}_{2}+Mc^{2})\gamma_{\nu}\big],
\end{split}
\end{equation}
where $\lambda_i$ and $\lambda_f$ are the helicity states respectively of initial and final electrons, and $S_i$ and $S_f$ are their 4-vectors spin. Now, the DCS for polarized electrons depends, in addition to other parameters, on $\lambda_i$ and $\lambda_f$. Therefore, if a flip of the electron spin occurs during the process, which means that:
\begin{equation}
\lambda_f=-\lambda_i=\pm1,~~~\Longrightarrow~~~\lambda_i\lambda_f=-1,
\end{equation}
the obtained DCS is called spin-flip DCS. In the opposite case where no spin flip occurs ($\lambda_f=\lambda_i=\pm1,~\Rightarrow~\lambda_i\lambda_f=1$), it is called spin non-flip DCS. One of the most important consistency tests to check is that the sum of the spin-flip and the spin non-flip DCSs must give the unpolarized DCS given in Eq.~(\ref{DCS_free})
\begin{equation}
\bigg(\frac{d\overline{\sigma}}{d\Omega_{f}}\bigg)_{\text{unpolarized}}^{\text{laser-free}}=\bigg(\frac{d\overline{\sigma}}{d\Omega_{f}}\bigg)_{\text{flip}}+\bigg(\frac{d\overline{\sigma}}{d\Omega_{f}}\bigg)_{\text{non-flip}}.
\end{equation}
Another important quantity to be introduced here is the degree of polarization (DP), which is defined as:
\begin{eqnarray}\label{DP}
\text{DP}=\frac{\big(d\overline{\sigma}/d\Omega_{f}\big)_{\text{non-flip}}-\big(d\overline{\sigma}/d\Omega_{f}\big)_{\text{flip}}}{\big(d\overline{\sigma}/d\Omega_{f}\big)_{\text{non-flip}}+\big(d\overline{\sigma}/d\Omega_{f}\big)_{\text{flip}}}.
\end{eqnarray} 
\subsection{Laser-assisted differential cross section for polarized electrons}
The monochromatic laser field of circular polarization is described by the following classical 4-potential:
\begin{equation}\label{potentiel_laser}
A_{\text{laser}}^{\mu}(\phi)=a^{\mu}_{1}\cos(\phi)+a^{\mu}_{2}\sin(\phi),
\end{equation}
where $a_{1}^{\mu}=(0,|\vec{a}|,0,0)$ and $a_{2}^{\mu}=(0,0,|\vec{a}|,0)$ are the orthogonal polarization 4-vectors satisfying the following relation: $a_{1}^{2}=a_{2}^{2}=a^{2}=-|\vec{a}|^{2}=(c\varepsilon_{0}/\omega)^{2}$, where $\varepsilon_{0}$ is the electric field strength and $\omega$ is the electromagnetic field frequency. $\phi=k.x$ is the phase of the electromagnetic field of wave 4-vector $k=(\omega/c,\vec{k})$ (with $k^{2}=0$).
To describe an electron dressed by an electromagnetic field, we use the solution of the following Dirac equation in the presence of the laser field:
\begin{eqnarray}\label{Dirac_Volkov}
\Big[\Big(\hat{p}-\frac{A}{c}\Big)^{2}-c^{2}-\frac{i}{2c}F_{\mu\nu}\sigma^{\mu\nu}\Big]\psi(x)=0,
\end{eqnarray}
where $F_{\mu\nu}=\partial_{\mu}A_{\nu}-\partial_{\nu}A_{\mu}$ represents the electromagnetic tensor and $\sigma^{\mu \nu}=\tfrac{1}{2}[\gamma^{\mu},\gamma^{\nu}]$. The general solution of Eq.~(\ref{Dirac_Volkov}) is an exact Dirac-Volkov plane wave function expressed as follows \cite{volkov}:
\begin{eqnarray}\label{volkov-function}
\psi_{p_{n}}(x)=\bigg(1+\frac{\slashed{k}\slashed{A}}{2c(k.p_{n})}\bigg)\frac{u(p_{n},s_{n})}{\sqrt{2Q_{n}V}}e^{iS(q_{n},x)},
\end{eqnarray}
where $u(p_{n},s_{n})$ is the free Dirac bispinor of the electron, which verifies the following relation: 
\begin{eqnarray}
\bar{u}(p_{n},s_{n})u(p_{n},s_{n})=2c^{2}.
\end{eqnarray}
The dressed electron is of 4-momentum $q_{n}=(Q_{n}/c,\vec{q}_{n})$ given by:
\begin{eqnarray}
q_{n}=p_{n}-\frac{a^{2}}{2c^{2}(k.p_{n})}k,\quad \text{with}~~\,q_{n}^{2}={m^{*}}^{2}c^{2},
\end{eqnarray}
where $m^{*}=\sqrt{1-a^{2}/c^{4}}$ is the effective mass of the electron in the presence of the laser field. The classical action $S(q_{n},x)$ in Eq.~(\ref{volkov-function}) has the following expression:
\begin{eqnarray}
S(q_{n},x)=-q_{n}.x-\frac{a_{1}.p_{n}}{c(k.p_{n})}\sin(\phi)+\frac{a_{2}.p_{n}}{c(k.p_{n})}\cos(\phi).
\end{eqnarray}
In the first Born approximation, using Feynman's rules and after a simple treatment, we express $S_{fi}$ in the presence of an electromagnetic field by:
\begin{eqnarray}
S_{fi}&=&\frac{-i}{\sqrt{16Q_{1}Q_{3}E_{2}E_{4}V^{4}}}\int d^{4}x~d^{4}y~\frac{d^{4}q}{(2\pi)^{4}}\frac{e^{i(q_{3}-q_{1}-q).x}e^{i(p_{4}-p_{2}+q).y}e^{-iz\sin(\phi-\phi_{0})}}{q^{2}+i\varepsilon}\nonumber \\
&\times& \bigg[\overline{u}(p_{3},s_{3})\bigg(C_{\mu}^{0}+C_{\mu}^{1}\cos(\phi)+C_{\mu}^{2}\sin(\phi)\bigg)u(p_{1},s_{1})\bigg]\bigg[\overline{u}(p_{4},s_{4})\gamma^{\mu}u(p_{2},s_{2})\bigg],
\end{eqnarray}
where
\begin{eqnarray}
z=\sqrt{\alpha_{1}^{2}+\alpha_{2}^{2}}~~~\text{and}~~~\phi_{0}=\arctan(\alpha_{2}/ \alpha_{1}),
\end{eqnarray}
with
\begin{eqnarray}
\alpha_{1}=\frac{a_{1}.p_{1}}{c(k.p_{1})}-\frac{a_{1}.p_{3}}{c(k.p_{3})}\quad\text{and}\quad\alpha_{2}=\frac{a_{2}.p_{1}}{c(k.p_{1})}-\frac{a_{2}.p_{3}}{c(k.p_{3})}.
\end{eqnarray}
The terms $C^{0}_{\mu}$, $C^{1}_{\mu}$ and $C^{2}_{\mu}$ are given by:
\begin{eqnarray}
\left\lbrace\begin{array}{ccc} 
C^{0}_{\mu}&=&\gamma_{\mu}-2k_{\mu}a^{2}\slashed{k}C(p_{1})C(p_{3}),\\
C^{1}_{\mu}&=&C(p_{1})\gamma_{\mu}\slashed{k}\slashed{a}_{1}+C(p_{3})\slashed{a}_{1}\slashed{k}\gamma_{\mu},\\
C^{2}_{\mu}&=&C(p_{1})\gamma_{\mu}\slashed{k}\slashed{a}_{2}+C(p_{3})\slashed{a}_{2}\slashed{k}\gamma_{\mu}.
\end{array}\right.\qquad\text{where}\quad C(p_{i})=\frac{1}{2c(k.p_{i})}.
\end{eqnarray}
Then, by integrating with respect to $d^{4}x$, $d^{4}y$ and $d^{4}q$, we obtain:
\begin{eqnarray}
S_{fi}&=&\frac{-i}{\sqrt{16Q_{1}Q_{3}E_{2}E_{4}V^{4}}}\sum_{\ell=-\infty}^{+\infty} \frac{(2\pi)^{4}\delta^{4}(p_{4}-p_{2}+q_{3}-q_{1}-\ell k)}{q^{2}+i\varepsilon}\nonumber \\
&\times& \bigg[\overline{u}(p_{3},s_{3})\bigg(C_{\mu}^{0}B_{\ell}(z)+C_{\mu}^{1}B_{1\ell}(z)+C_{\mu}^{2}B_{2\ell}(z)\bigg)u(p_{1},s_{1})\bigg]\bigg[\overline{u}(p_{4},s_{4})\gamma^{\mu}u(p_{2},s_{2})\bigg],
\end{eqnarray}
where $\ell$ is the number of photons exchanged and $q=q_{3}-q_{1}-\ell k$ is the 4-momentum transfer. The functions $B_{\ell}$, $B_{1\ell}$ and $B_{2\ell}$ are obtained using Fourier transforms into Bessel functions \cite{Abramowitz}, such that:
\begin{eqnarray}
\Big(1,\cos(\phi),\sin(\phi)\Big)e^{-iz\sin(\phi-\phi_{0})}=\sum_{\ell=-\infty}^{+\infty}\Big(B_{\ell}(z),B_{1\ell}(z),B_{2\ell}(z)\Big),
\end{eqnarray}
with
\begin{align}
\begin{split}
\begin{bmatrix}
B_{\ell}(z)\\
B_{1\ell}(z)\\
B_{2\ell}(z) \end{bmatrix}=\begin{bmatrix}J_{\ell}(z)e^{i\ell\phi_{0}}\\
\big(J_{\ell+1}(z)e^{i(\ell+1)\phi_{0}}+J_{\ell-1}(z)e^{i(\ell-1)\phi_{0}}\big)/2\\
\big(J_{\ell+1}(z)e^{i(\ell+1)\phi_{0}}-J_{\ell-1}(z)e^{i(\ell-1)\phi_{0}}\big)/2i
 \end{bmatrix}.
\end{split}
\end{align}
As in the laser-free case of the DCS (\ref{DCS_0}), we give the DCS in the presence of an external laser field by the following formula:
\begin{eqnarray}\label{DCS_laser_0}
d\bar{\sigma}=\sum_{\ell=-\infty}^{+\infty} \frac{Vd^{3}\vec{q}_{3}}{(2\pi)^{3}}\frac{Vd^{3}\vec{p}_{4}}{(2\pi)^{3}}\frac{1}{|J_{\text{inc.}}|T}\frac{(2\pi)^{4}VT\delta^{4}(p_{4}-p_{2}+q_{3}-q_{1}-\ell k)}{16Q_{1}E_{2}Q_{3}E_{4}V^{4}q^{4}}|\overline{\mathcal{M}_{fi}^{\ell}}|^{2},
\end{eqnarray}
where $|J_{\text{inc.}}|=[|\vec{q}_{1}|c^{2}]/[VQ_{1}]$ is the flux of incident electrons dressed by the laser field. The associated scattering amplitude $|\overline{\mathcal{M}_{fi}^{\ell}}|$, in case of well defined spin electrons, will take the following form:
\begin{equation}\label{trace3}
\begin{split}
|\overline{\mathcal{M}_{fi}^{\ell}}|^{2}=&\text{Tr}\big[(c\slashed{p}_{3}+c^{2})\Gamma_{\mu}^{\ell}(\frac{1+\lambda_{1}\gamma_{5}\slashed{S}_{1}}{2})(c\slashed{p}_{1}+c^{2})\Gamma_{\nu}^{\ell\dagger}(\frac{1+\lambda_{3}\gamma_{5}\slashed{S}_{3}}{2})\big] \\
& \times \frac{1}{2} \text{Tr}\big[(c\slashed{p}_{4}+Mc^{2})\gamma^{\mu}(c\slashed{p}_{2}+Mc^{2})\gamma^{\nu}\big],
\end{split}
\end{equation}
with $\Gamma_{\nu}^{\ell\dagger}$ is the conjugate of $\Gamma_{\mu}^{\ell}$ defined by:
\begin{eqnarray}
\Gamma_{\mu}^{\ell}=C_{\mu}^{0}B_{\ell}(z)+C_{\mu}^{1}B_{1\ell}(z)+C_{\mu}^{2}B_{2\ell}(z).
\end{eqnarray}
Finally, the DCS expression in the presence of an electromagnetic wave and for well-defined spin states of the electron is given by:
\begin{eqnarray}\label{DCS_with_laser_and_spin_polarisation}
\bigg(\frac{d\overline{\sigma}}{d\Omega_{f}}\bigg)^{\text{laser-assisted}}_{\text{$e$-polarized}}(\lambda_{1},\lambda_{3})=\sum_{\ell=-\infty}^{+\infty}\frac{d\overline{\sigma}^{\ell}}{d\Omega_{f}}(\lambda_{1},\lambda_{3})=\frac{1}{16(2\pi)^{2}Mc^{6}q^{4}}\sum_{\ell=-\infty}^{+\infty}\frac{|\vec{q}_{3}|}{|\vec{q}_{1}|}\frac{|\overline{\mathcal{M}^{\ell}_{fi}}|^{2}}{|g'(Q_{3})|},
\end{eqnarray}
where the function $g'(Q_{3})$ is obtained by using the property in Eq.~(\ref{property_1}): 
\begin{eqnarray}
g'(Q_{3})=\bigg[\frac{Q_{1}}{c^{2}}+M+\frac{\ell\omega}{c^{2}}\bigg]-\frac{Q_{3}}{c^{2}|\vec{q_{3}}|}\bigg[|\vec{q}_{1}|F(\theta_{i},\theta_{f},\varphi_{i},\varphi_{f})+\frac{\ell\omega}{c^{2}}\cos(\theta_{f})\bigg].
\end{eqnarray}
For the spin-flip and spin non-flip laser-assisted DCSs, we have:
\begin{equation}
\begin{split}
\bigg(\frac{d\overline{\sigma}}{d\Omega_{f}}\bigg)_{\text{flip}}&=\bigg(\frac{d\overline{\sigma}}{d\Omega_{f}}\bigg)^{\text{laser-assisted}}_{\text{$e$-polarized}}(\lambda_{1}=-\lambda_{3}=\pm1),\\
\bigg(\frac{d\overline{\sigma}}{d\Omega_{f}}\bigg)_{\text{non-flip}}&=\bigg(\frac{d\overline{\sigma}}{d\Omega_{f}}\bigg)^{\text{laser-assisted}}_{\text{$e$-polarized}}(\lambda_{1}=\lambda_{3}=\pm1).
\end{split}
\end{equation}
The sum of the two (spin-flip and non-flip DCSs) should again yield the unpolarized laser-assisted DCS that was previously calculated in a recent work \cite{Dahiri_2021}. \\
The calculation of the traces appearing in Eqs.~(\ref{trace1}), (\ref{trace2}) and (\ref{trace3}) is performed with the help of the symbolic-algebra program Feyncalc \cite{feyncalc}. We mention here that the result obtained is too long and cumbersome to be included here due to the complexity of the calculations. For example, the result of the trace calculation in the presence of the laser (given in eq.~(\ref{trace3})) yielded about 63 different antisymmetric tensors,  $\epsilon(a,b,c,d)=\epsilon_{\mu\nu\rho\sigma}a^{\mu}b^{\nu}c^{\rho}d^{\sigma}$ for all four-vectors $a, b, c$ and $d$, which appear when four matrices $\gamma$ meet the fifth matrix $\gamma_{5}$ inside the trace $(\text{Tr}[\gamma^{\mu}\gamma^{\nu}\gamma^{\rho}\gamma^{\sigma}\gamma^{5}]=-4i\epsilon^{\mu\nu\rho\sigma})$. These resulting tensors were calculated analytically and replaced.
\section{Results and discussion}\label{Sec2}
After this detailed theoretical treatment of elastic e-p scattering in the presence of a laser field and for electrons of well-defined spin polarization, we will now present the different numerical results of this work. The geometry that has been found to give good results is one with spherical angles of $\theta_{i}=\varphi_{i}=15^{\circ}$ and $\varphi_{f}=105^{\circ}$. The final angle $\theta_f$ is varied from $-180^{\circ}$ to $180^{\circ}$ to reveal the angular distribution of the DCS. The incident elecron kinetic energy is (unless otherwise stated) $T_i=2.7$ keV. The direction of the electromagnetic field is that represented by the wave vector $\vec{k}$; it is chosen along the axis ($Oz$). We start with the first result in which the element $d\overline{\sigma}^{\ell}/d\Omega_{f}(\lambda_{1},\lambda_{3})$ of Eq.~(\ref{DCS_with_laser_and_spin_polarisation}) represents the DCS of each number of photons exchanged.
\begin{figure}[H]
\centering
  \begin{minipage}[t]{0.43\textwidth}
  \centering
    \includegraphics[width=\textwidth]{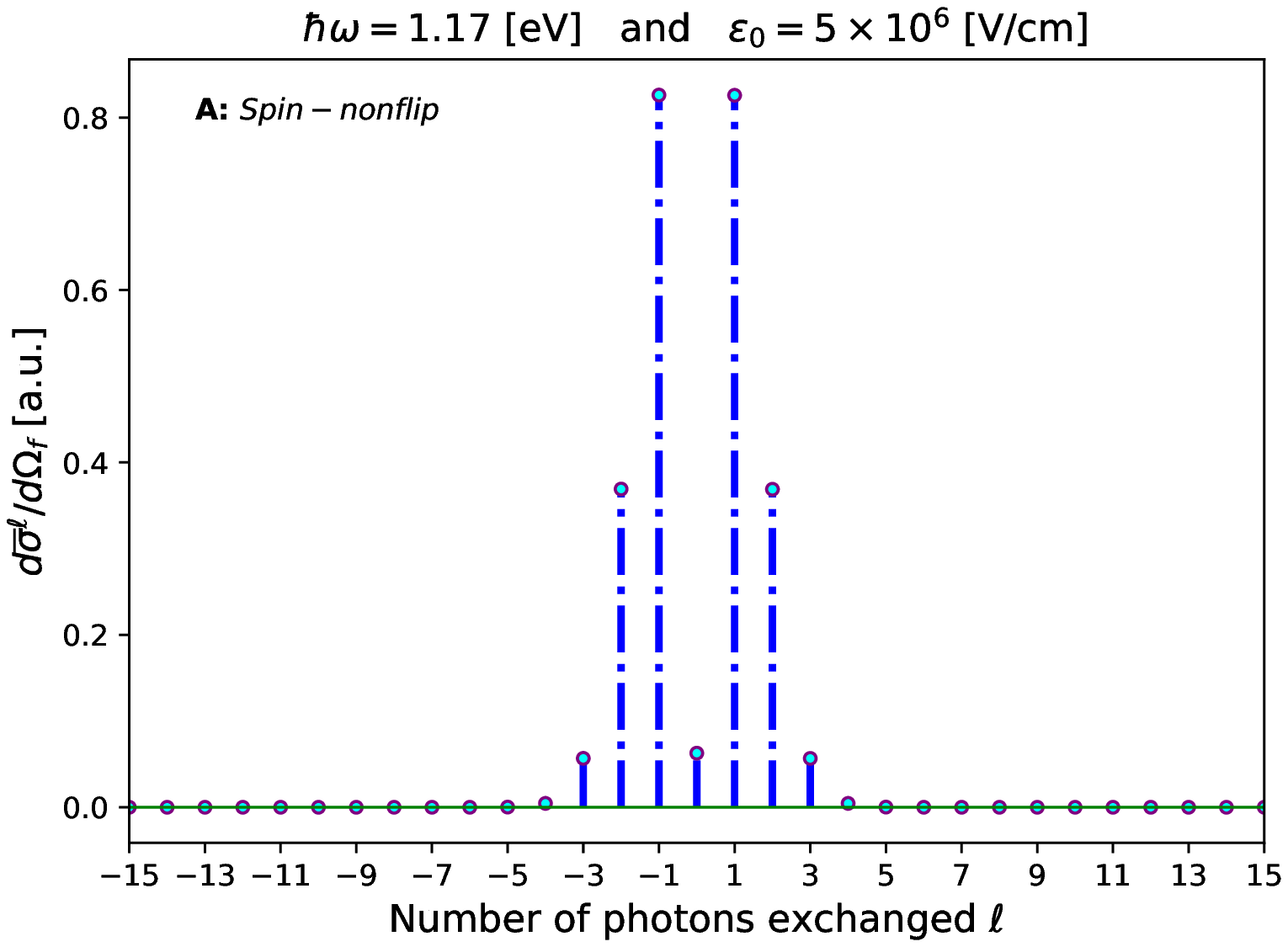}
  \end{minipage}
  \hspace*{0.25cm}
  \begin{minipage}[t]{0.43\textwidth}
  \centering
    \includegraphics[width=\textwidth]{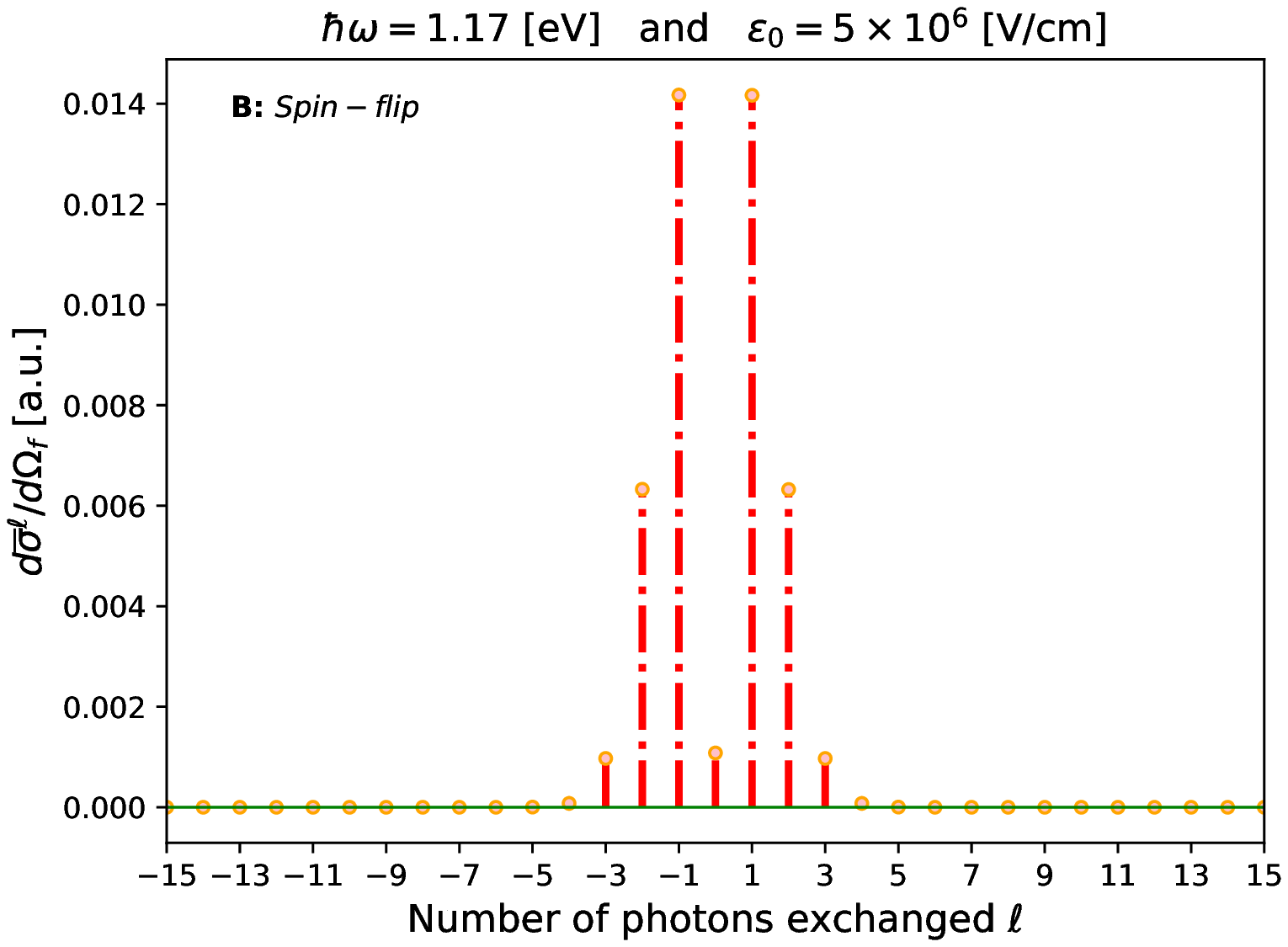}
  \end{minipage}
  \begin{minipage}[t]{0.43\textwidth}
  \centering
    \includegraphics[width=\textwidth]{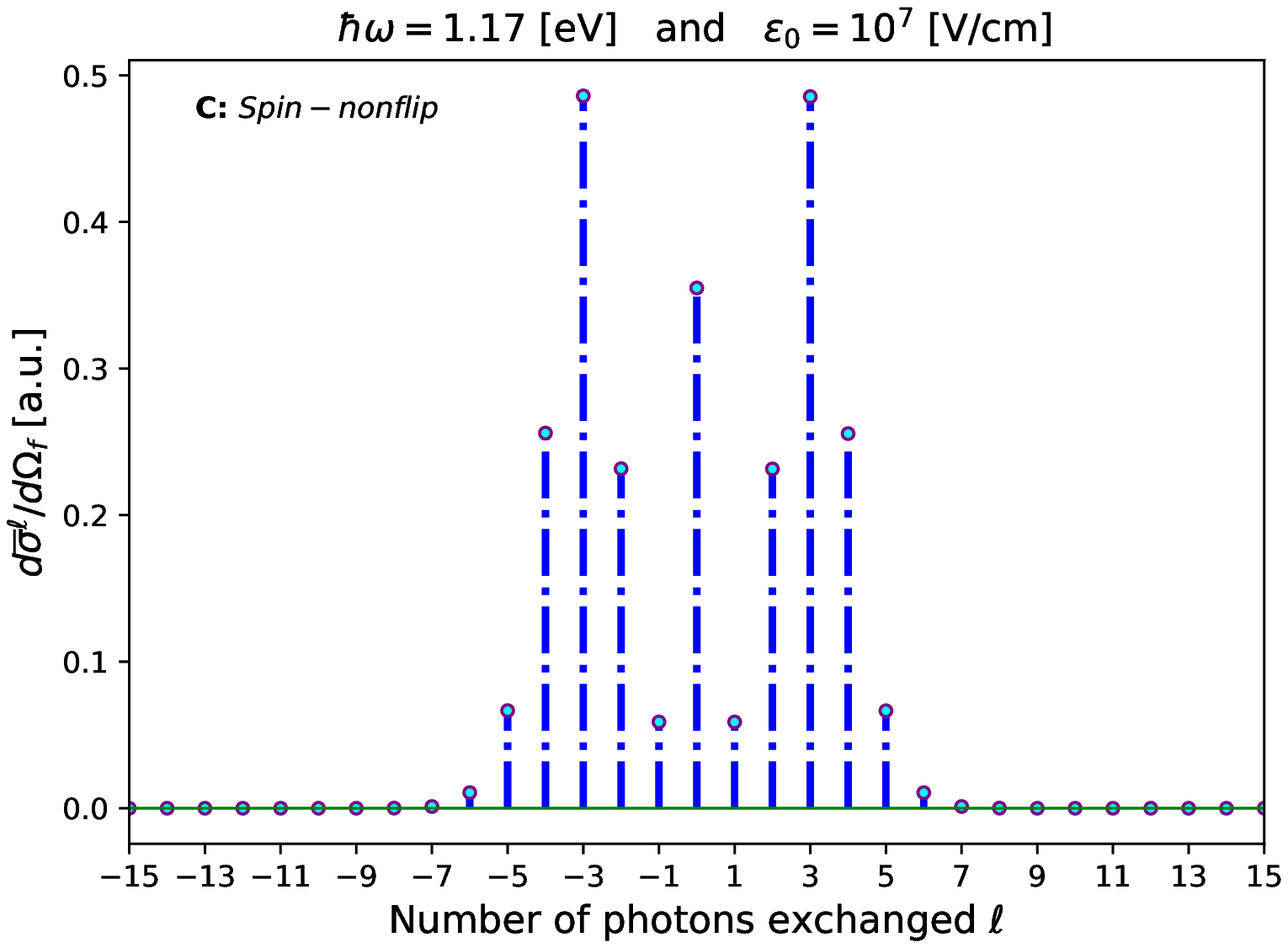}
  \end{minipage}
  \hspace*{0.25cm}
  \begin{minipage}[t]{0.43\textwidth}
  \centering
    \includegraphics[width=\textwidth]{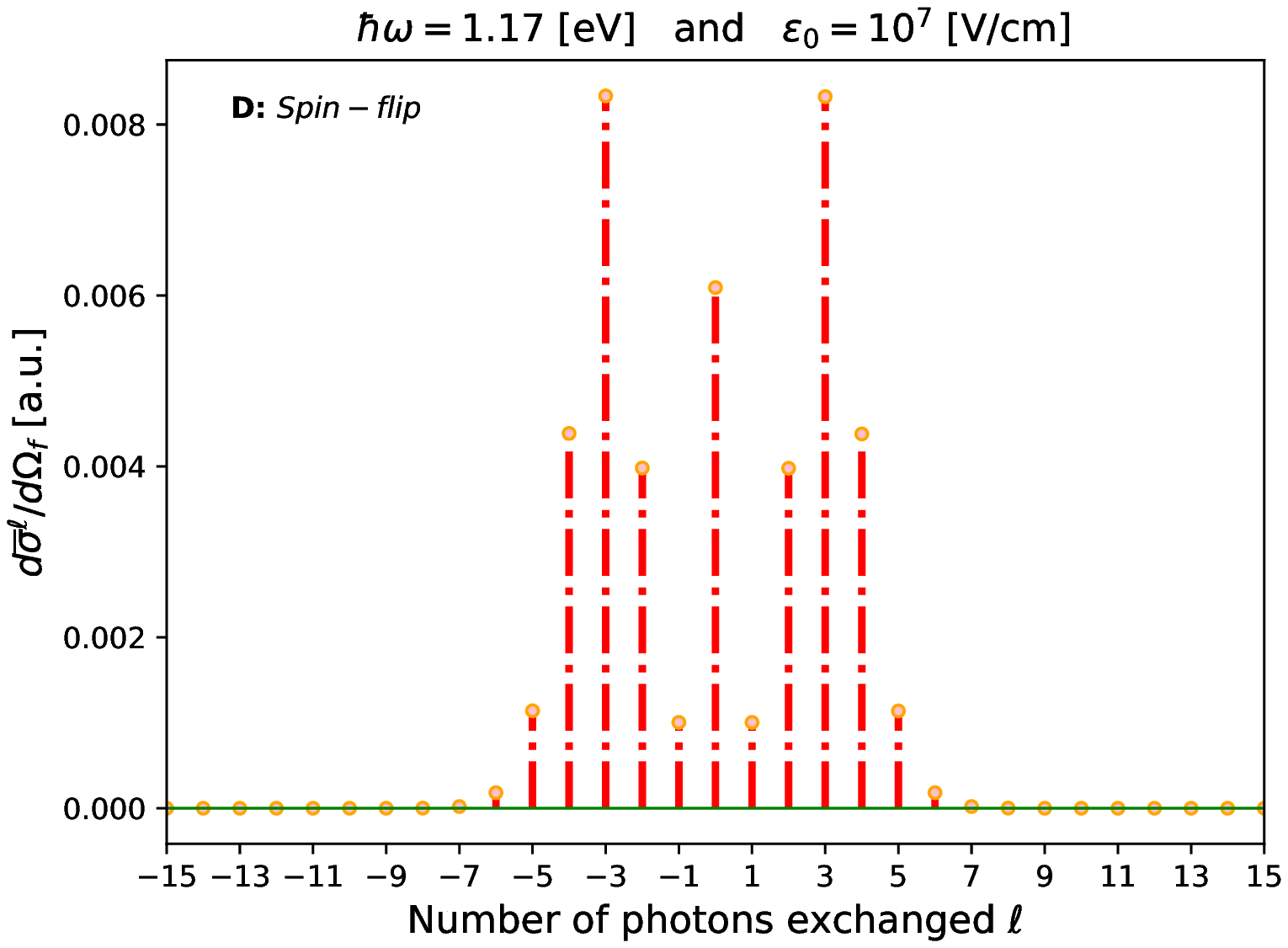}
  \end{minipage}
  \caption{Differential cross section $d\overline{\sigma}^{\ell}/d\Omega_{f}(\lambda_{1},\lambda_{3})$ for spin-flip ($\lambda_{1}=-\lambda_{3}=1$) and spin non-flip ($\lambda_{1}=\lambda_{3}=1$) as a function of the number of photons exchanged $\ell$. The scattering final angle is $\theta_{f}=0^{\circ}$. The field strength $\varepsilon_0$ and frequency $\hbar\omega$ are indicated above each figure.} \label{Figure:1}
\end{figure}
Fig.~\ref{Figure:1} gives information about the number of possible photons to be exchanged during the scattering process, treating the two cases of polarization of spins initially defined in spin Up ($\lambda_{1}=1$). For a laser field frequency $\hbar\omega=1.17$ eV and strength $\varepsilon_{0}=5\times10^{6}$ V/cm, the DCS, $d\overline{\sigma}^{\ell}/d\Omega_{f}(\lambda_{1},\lambda_{3})$, in the two Figs.~\ref{Figure:1}A and \ref{Figure:1}B for spin non-flip and spin-flip varies in magnitude for the same number of photons exchanged $-3\leq \ell \leq3$. This result is also seen by comparing Fig.~\ref{Figure:1}C with Fig.~\ref{Figure:1}D when the strength of the same laser field increases to $\varepsilon_{0}=10^{7}$ V/cm, but now the number of photons exchanged $\ell$ varies from $-5$ to $5$. We notice that by increasing the field strength the number of possible exchanged photons becomes larger. This number of photons exchanged $\ell$ corresponds to non-zero values of DCS, while the DCS becomes zero from some values higher than $\ell$ in absolute value (the number of exchanged photons $\ell$ depends strongly on the laser field strength). This value from which the DCS cancels is called the cutoff of the multiphoton scattering process. For a value $|\ell|$ greater than the cutoff, the summed laser-assisted DCS (\ref{DCS_with_laser_and_spin_polarisation}) becomes equal to its corresponding one in the absence of the field. Then, we say that the laser assisted process verifies the sum rule formulated as follows: 
\begin{equation}\label{sumrule}
\frac{d\overline{\sigma}_{free}}{d\Omega_{f}}(\lambda_{1},\lambda_{3})=\sum_{\ell=-\text{cutoff}}^{+\text{cutoff}}\frac{d\overline{\sigma}^{\ell}}{d\Omega_{f}}(\lambda_{1},\lambda_{3}).
\end{equation}
\begin{figure}[H]
\centering
  \begin{minipage}[h]{0.43\textwidth}
  \centering
    \includegraphics[width=\textwidth]{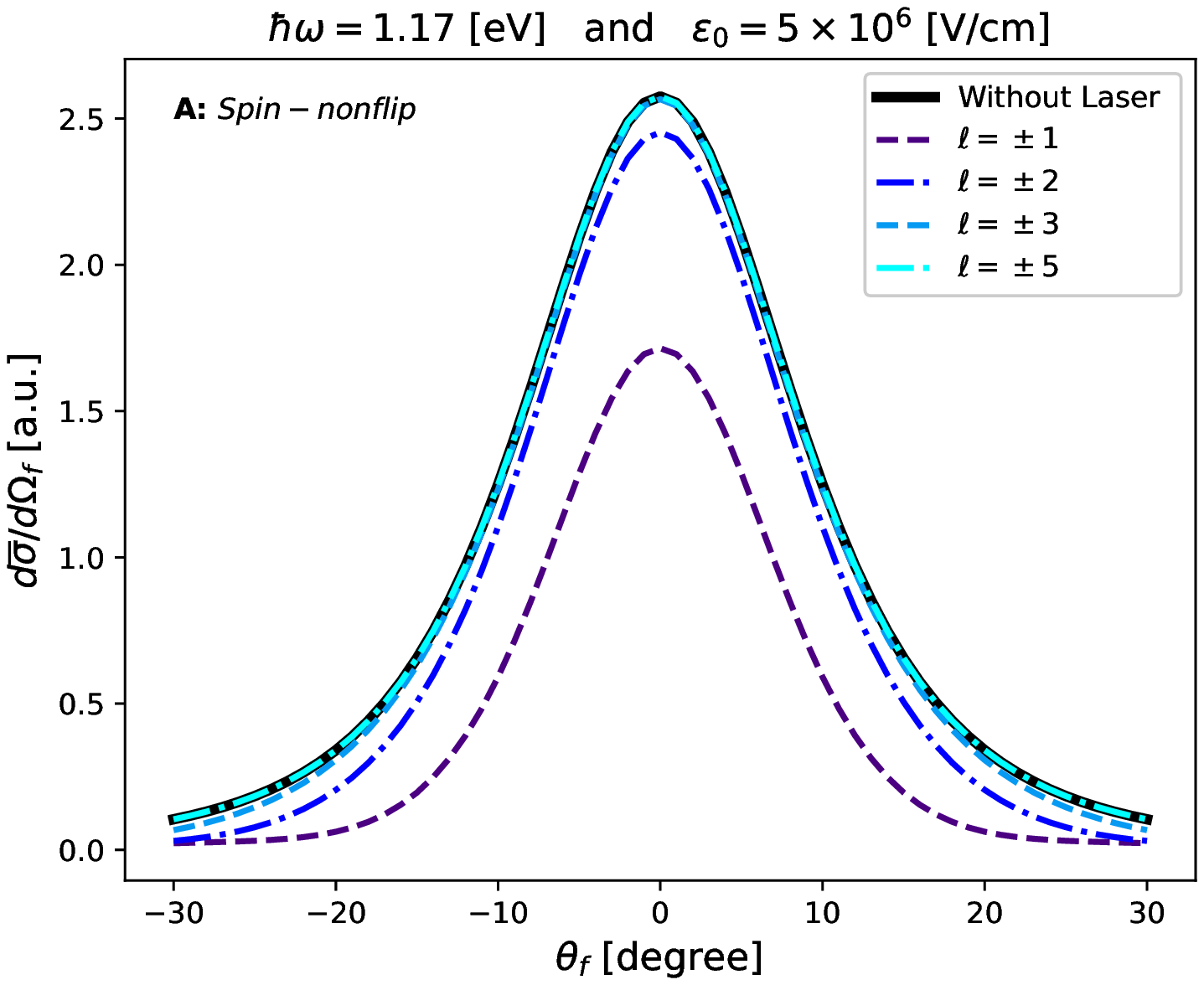}
  \end{minipage}
  \hspace*{0.25cm}
  \begin{minipage}[h]{0.43\textwidth}
  \centering
    \includegraphics[width=\textwidth]{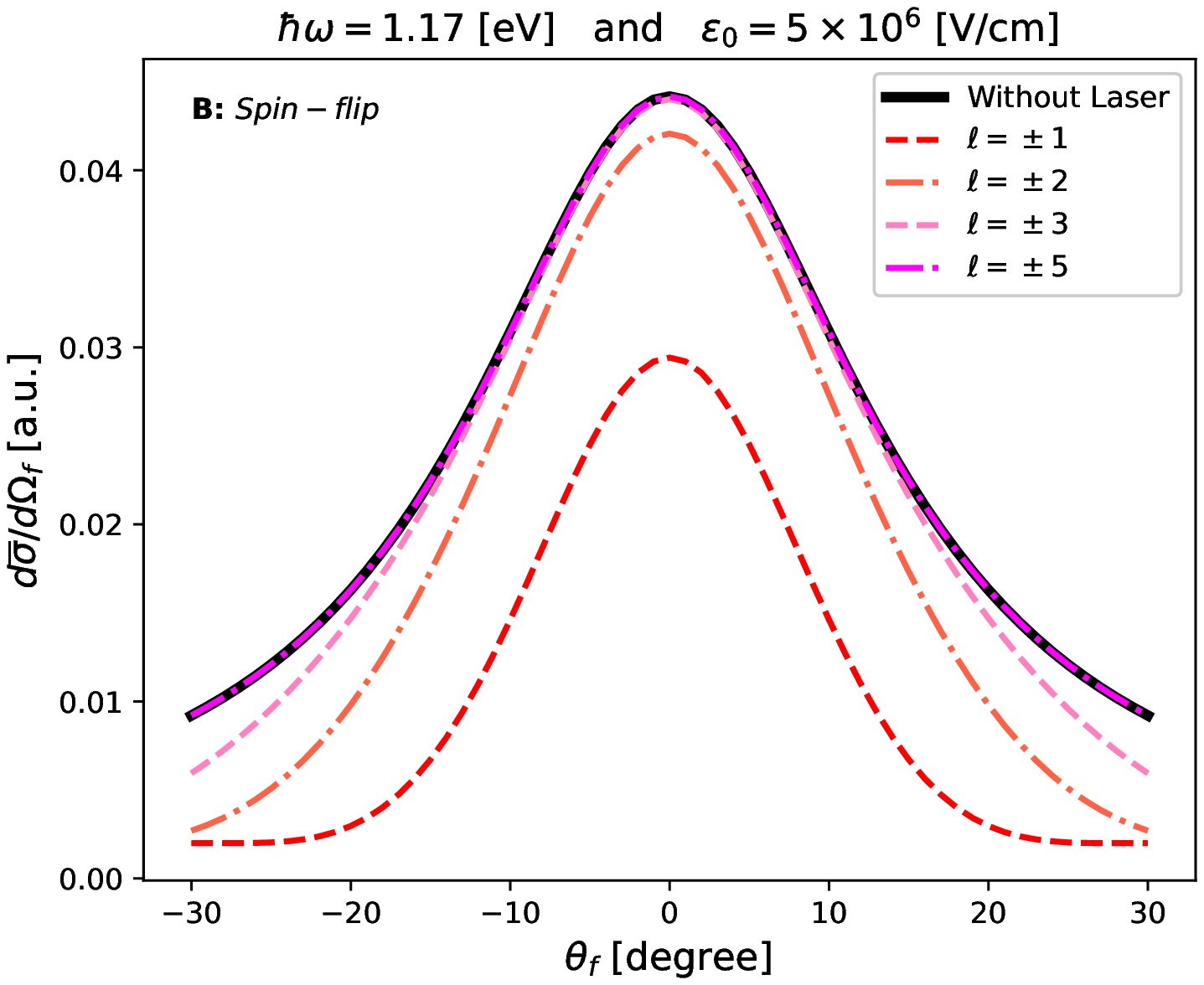}
  \end{minipage}
  \begin{minipage}[h]{0.43\textwidth}
  \centering
    \includegraphics[width=\textwidth]{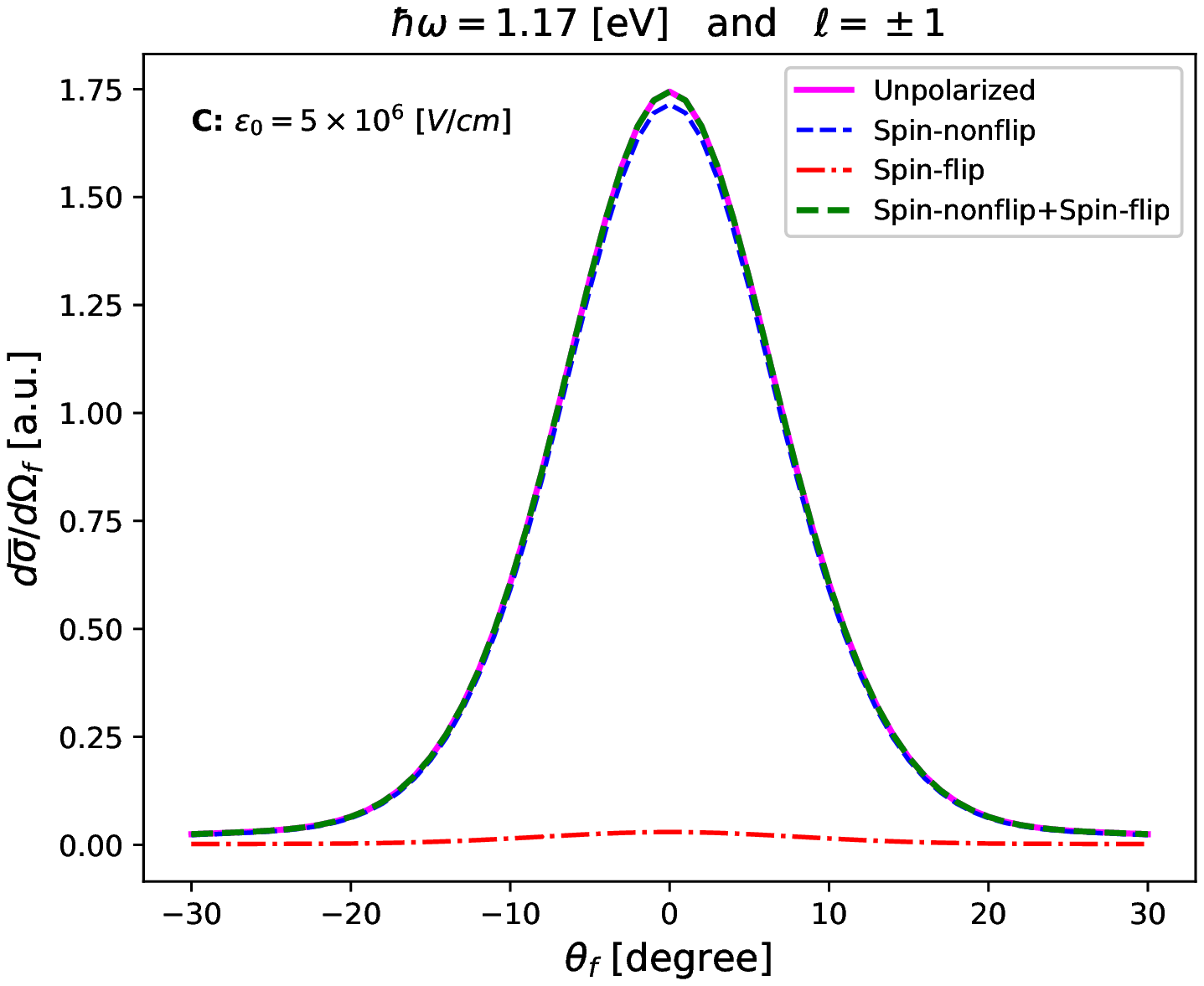}
  \end{minipage}
  \hspace*{0.25cm}
  \begin{minipage}[h]{0.43\textwidth}
  \centering
    \includegraphics[width=\textwidth]{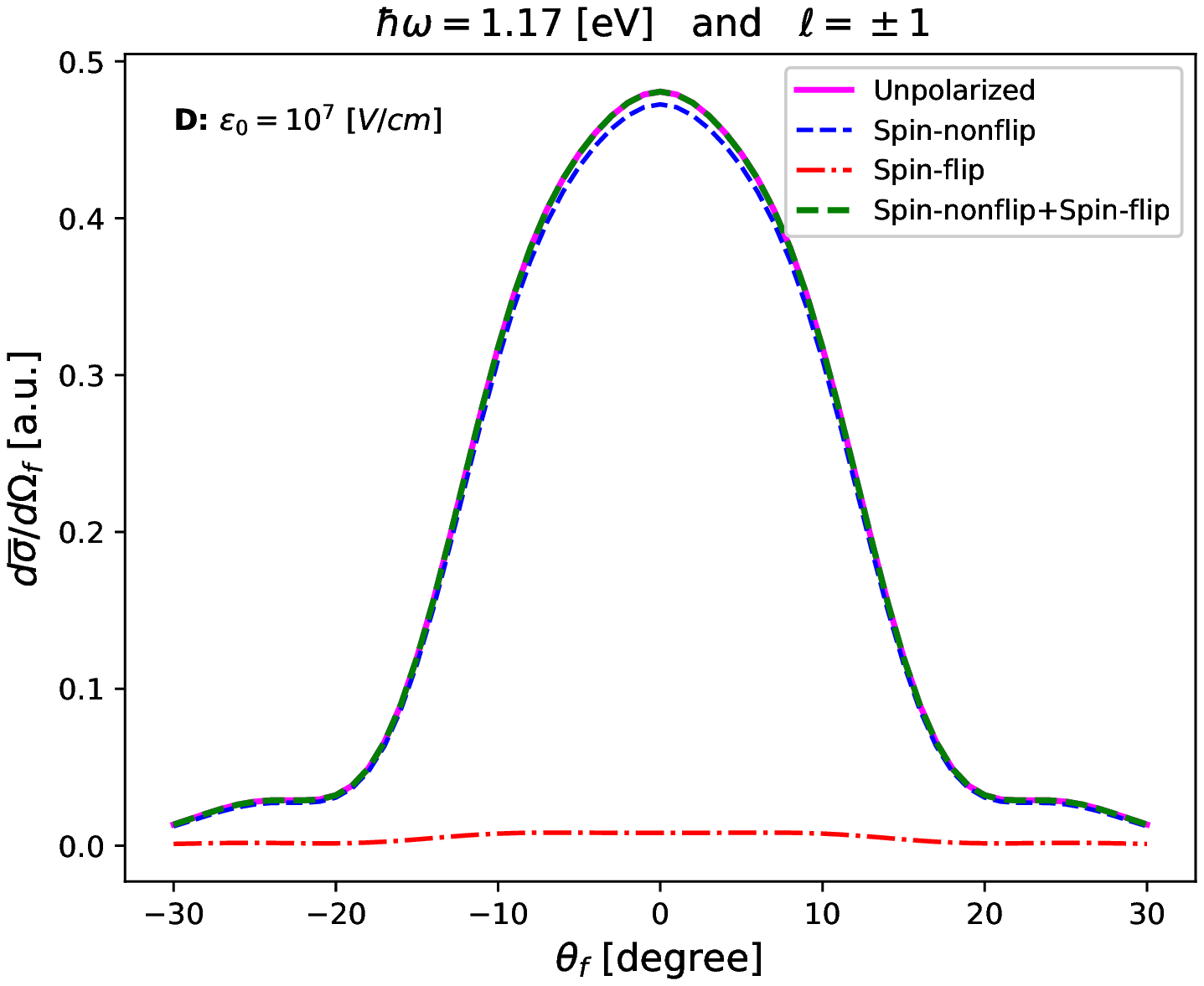}
  \end{minipage}
  \caption{Variation of the summed laser-assisted DCS (\ref{DCS_with_laser_and_spin_polarisation}) for polarized electrons as a function of the scattering angle $\theta_{f}$ for different summations over $\ell$. The notation $\ell=\pm i$ means that we have summed over $-i\leq\ell\leq+i$.} \label{Figure:2}
\end{figure}
The variation of the photon number $\ell$ given in the final expression of DCS (eq.~(\ref{DCS_with_laser_and_spin_polarisation})), which was introduced by Bessel functions of the first order, was a subject of observation of the absorption and free multiphoton emission in elastic electron-atom collisions in the presence of a cw-CO$_{2}$ laser \cite{andrick} and in the case of pulsed CO$_{2}$ laser \cite{Weingartshofer_1977}. In similar studies, the experimental data of these laser-assisted elastic electron scattering processes are in reasonable agreement with the predictions of the Kroll-Watson theory \cite{kroll}. 
The results in Fig.~\ref{Figure:1} will be considered in the discussion of the effect of the circularly polarized laser field on the DCS in Fig.~\ref{Figure:2} by varying the scattering angle $\theta_{f}$.
In Figs.~\ref{Figure:2}A and \ref{Figure:2}B, we have plotted the summed laser-assisted DCS (\ref{DCS_with_laser_and_spin_polarisation}) respectively for spin non-flip and spin-flip of electrons, summing over different numbers of photons exchanged $\ell$ from $-i$ to $i$ (with $i=\{1,2,3,5\}$). For these two figures, we can see that the applied Nd:YAG laser field, of frequency $\hbar\omega=1.17$ eV, has a decreasing effect on the DCS. This diminution becomes more important in the case of a single photon exchange ($\ell=\pm1$). While by increasing the number of photons, the DCS increases and becomes closer to that without field. We notice, in these two figures, that when the multiphoton process exchanges $5$ photons ($\ell=\pm 5$), the DCS in the presence of the laser field becomes closer to the one without the laser field, which confirms the result of the sum rule (\ref{sumrule}) already found in Fig.~\ref{Figure:1}A and \ref{Figure:1}B. On the other hand and according to the results obtained so far, we note that the spin non-flip DCS ($\lambda_{1}=\lambda_{3}=1$) is higher than the spin-flip one ($\lambda_{1}=-\lambda_{3}=1$). This remark is clearly observed in Fig.~\ref{Figure:2}C by comparing the blue curve with the red one for a laser field strength $\varepsilon_{0}=5\times10^{6}$ V/cm. We moved on to Fig.~\ref{Figure:2}D, where we raised the strength to $\varepsilon_{0}=10^{7}$ V/cm, we still found the difference between these two DCSs (flip and non-flip) for $\ell=\pm1$. 
Figs.~\ref{Figure:2}C and \ref{Figure:2}D show that the sum of these two DCSs always coincides with the unpolarized DCS. To give a closer picture of the effect of the laser field, for another type of laser, on each spin polarization, we will introduce numerical results in table \ref{table1} by increasing the field strength up to the value $\varepsilon_{0}=10^{8}$ V/cm. 
\begin{table}[h]
  \centerline{%
    \begin{tabular}{l*6{S}}
      \toprule
      \midrule
      \multirow{2}{*}{$\varepsilon_{0}\, \text{[V/cm]}$} & \multicolumn{2}{l}{\:\:\:\:\qquad\qquad non-flip DCS($\lambda_{1}=\lambda_{3}=1$)~[a.u.]}  & \multicolumn{2}{l}{\:\:\:\:\qquad\qquad flip DCS($\lambda_{1}=-\lambda_{3}=1$)~[a.u.]} \\
      \cmidrule(r){2-3}
      \cmidrule(r){4-5}
      & {~~$\hbar\omega=0.117$ eV} & {~~$\hbar\omega=1.17$ eV} & {~~~\qquad$\hbar\omega=0.117$ eV} & {~~~\qquad$\hbar\omega=1.17$ eV}\\
      \midrule
      ~~$10$     & 2.57555 & 2.57555 & 4.417$\times10^{-2}$ & 4.417$\times10^{-2}$   \\
      ~~$10^{2}$ & 2.57555 & 2.57555 & 4.417$\times10^{-2}$ & 4.417$\times10^{-2}$   \\
      ~~$10^{3}$ & 2.57555 & 2.57555 & 4.417$\times10^{-2}$ & 4.417$\times10^{-2}$   \\
      ~~$10^{4}$ & 2.57302 & 2.57555 & 4.412$\times10^{-2}$ & 4.417$\times10^{-2}$   \\
      ~~$10^{5}$ & 4.726$\times10^{-1}$ & 2.57555 & 8.105$\times10^{-3}$ & 4.417$\times10^{-2}$  \\
      ~~$10^{6}$ & 5.722$\times10^{-2}$ & 2.57302 & 9.808$\times10^{-4}$ & 4.412$\times10^{-2}$   \\
      ~~$10^{7}$ & 7.697$\times10^{-3}$ & 4.726$\times10^{-1}$ & 1.319$\times10^{-4}$ & 8.101$\times10^{-3}$   \\
      ~~$10^{8}$ & 7.373$\times10^{-3}$ & 5.644$\times10^{-2}$ & 1.141$\times10^{-5}$ & 9.598$\times10^{-4}$   \\
      \midrule
      \bottomrule
    \end{tabular}
  }%
\caption{Numerical values of spin-flip and non-flip DCSs as a function of $\varepsilon_{0}$ for different frequencies and for summation over the $\ell$ number from $-1$ to $1$. The scattering angle $\theta_f=0^{\circ}$.}\label{table1}
\end{table}
In table~\ref{table1}, we give a numerical vision of the effect of the electromagnetic field strength on the e-p scattering process. We see that the effect of a Nd:YAG laser field ($\hbar\omega=1.17$ eV) begins from a strength $\varepsilon_{0}=10^{6}$ V/cm and is manifested by a decrease in both spin-flip and non-flip DCSs. When the laser field is pulsed CO$_{2}$ ($\hbar\omega=0.117$ eV), its effect on DCSs starts when the strength $\varepsilon_{0}=10^{4}$ V/cm. For low strengths ranging from $10$ to $10^{4}$ V/cm and for $\ell=\pm1$, we have noted that the DCS for both polarization cases takes constant values that correspond to those in the absence of the field, which verifies the sum rule for $\ell=\pm1$. In terms of spin, the non-flip DCS in the case where $\hbar\omega=0.117$ eV and the strength ranges from $10^{5}$ to $10^{8}$ V/cm has decreased slightly compared to the spin-flip DCS. In Fig.~\ref{Figure:3}, we plot the curves representing the values given in table \ref{table1}, with the addition of another laser type (He:Ne) of frequency 2 eV.  
We notice that the frequency of the laser field can change the variation of the DCS with respect to $ \varepsilon_{0} $ for both cases of spin polarization. We also observe that for the CO$_{2}$ laser, the effect of the DCS is very important. By comparing between the DCSs in the spin-flip and non-flip cases, both figures show that they have the same decrease by varying the strength $\varepsilon_{0}$. So, if there is a difference between these two polarizations, it will be small. From Fig.~\ref{Figure:3}, it appears that the laser field effect diminishes at higher frequencies.
\begin{figure}[ht]
\centering
  \begin{minipage}[t]{0.42\textwidth}
  \centering
    \includegraphics[width=\textwidth]{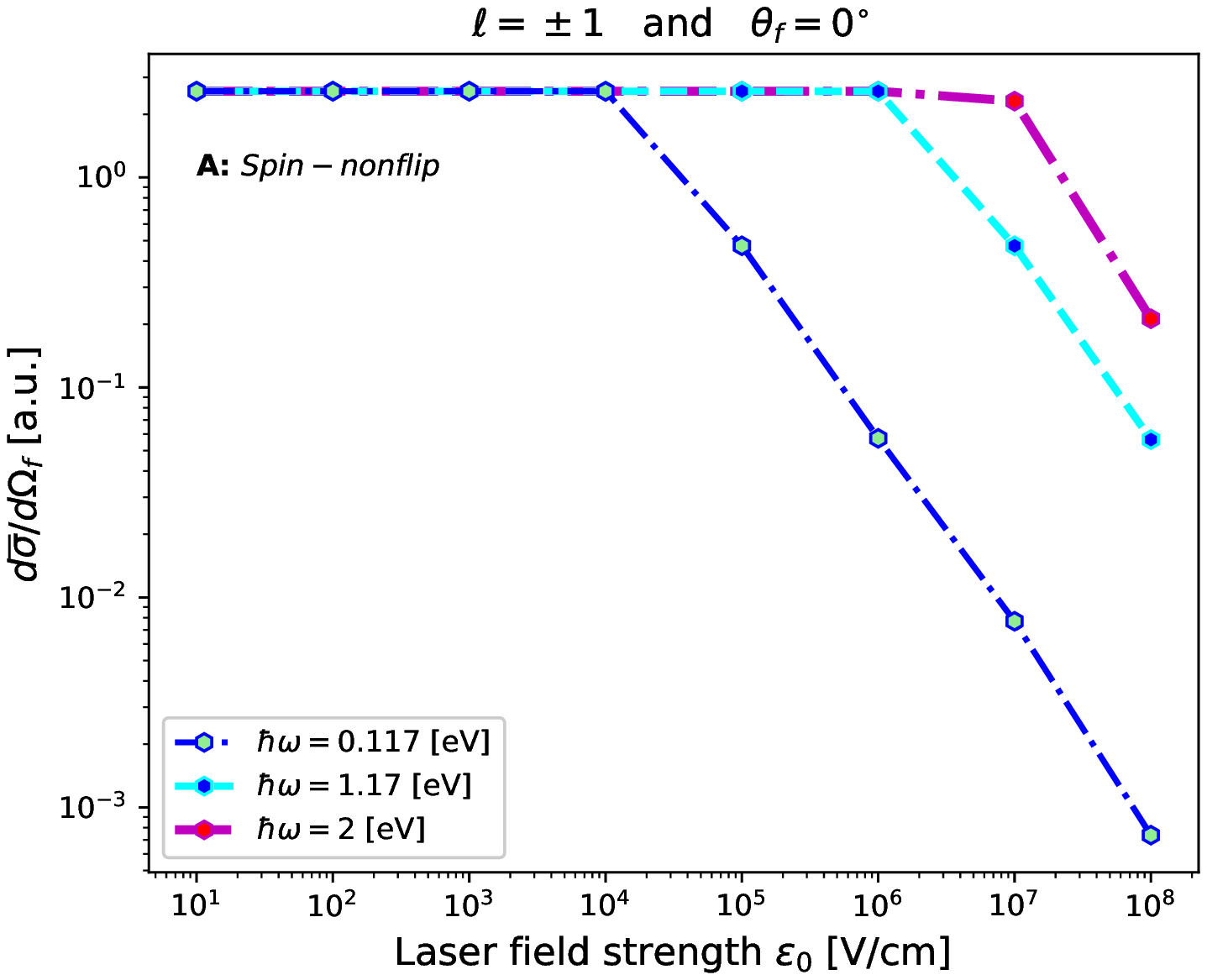}
  \end{minipage}
  \hspace*{0.25cm}
  \begin{minipage}[t]{0.42\textwidth}
  \centering
    \includegraphics[width=\textwidth]{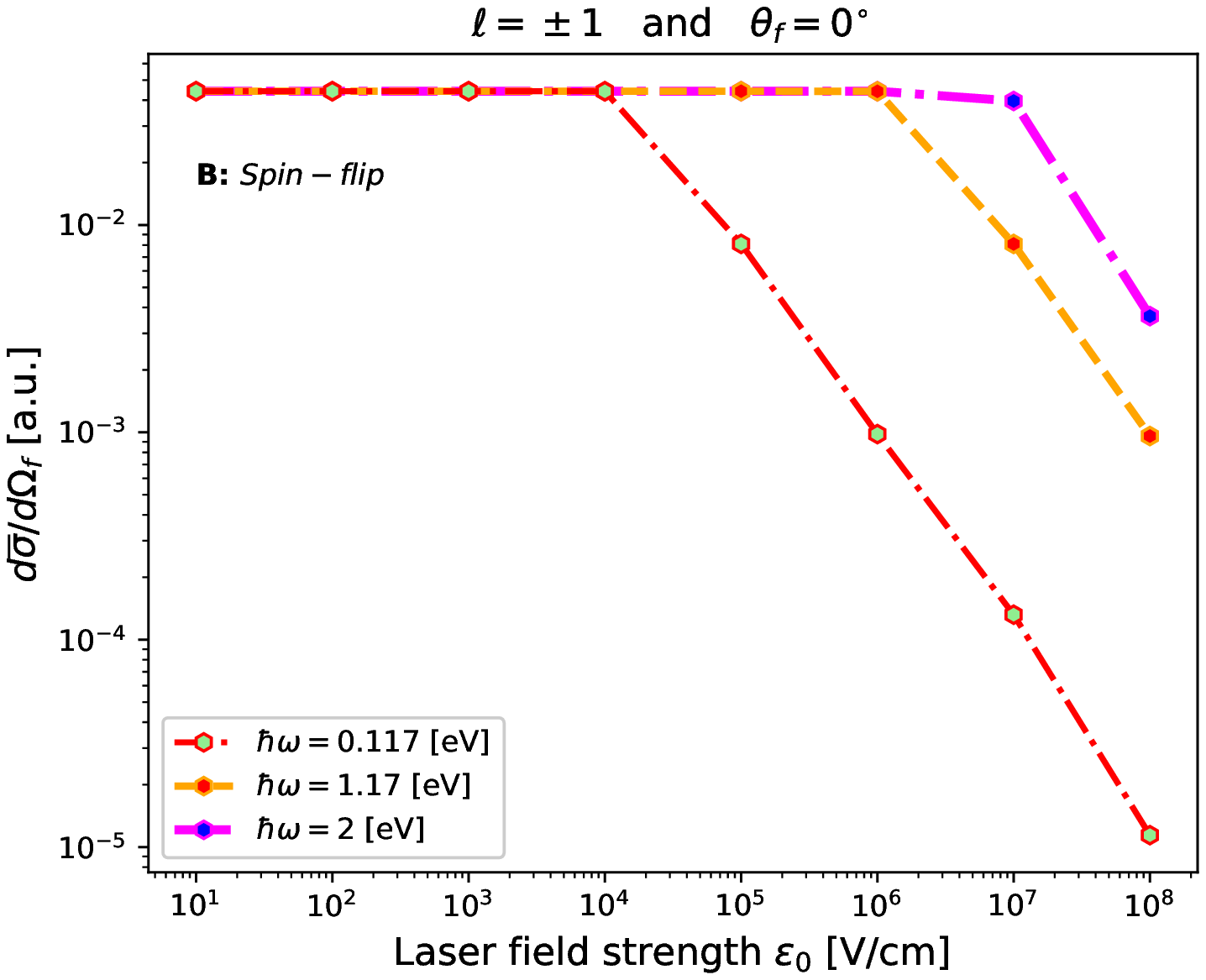}
  \end{minipage}
   \caption{Variation of (A) spin non-flip and (B) spin-flip DCSs as a function of the laser field strength for different laser frequencies.} \label{Figure:3}
\end{figure}
 \begin{figure}[ht]
  \centering
  \begin{minipage}[t]{0.42\textwidth}
  \centering
    \includegraphics[width=\textwidth]{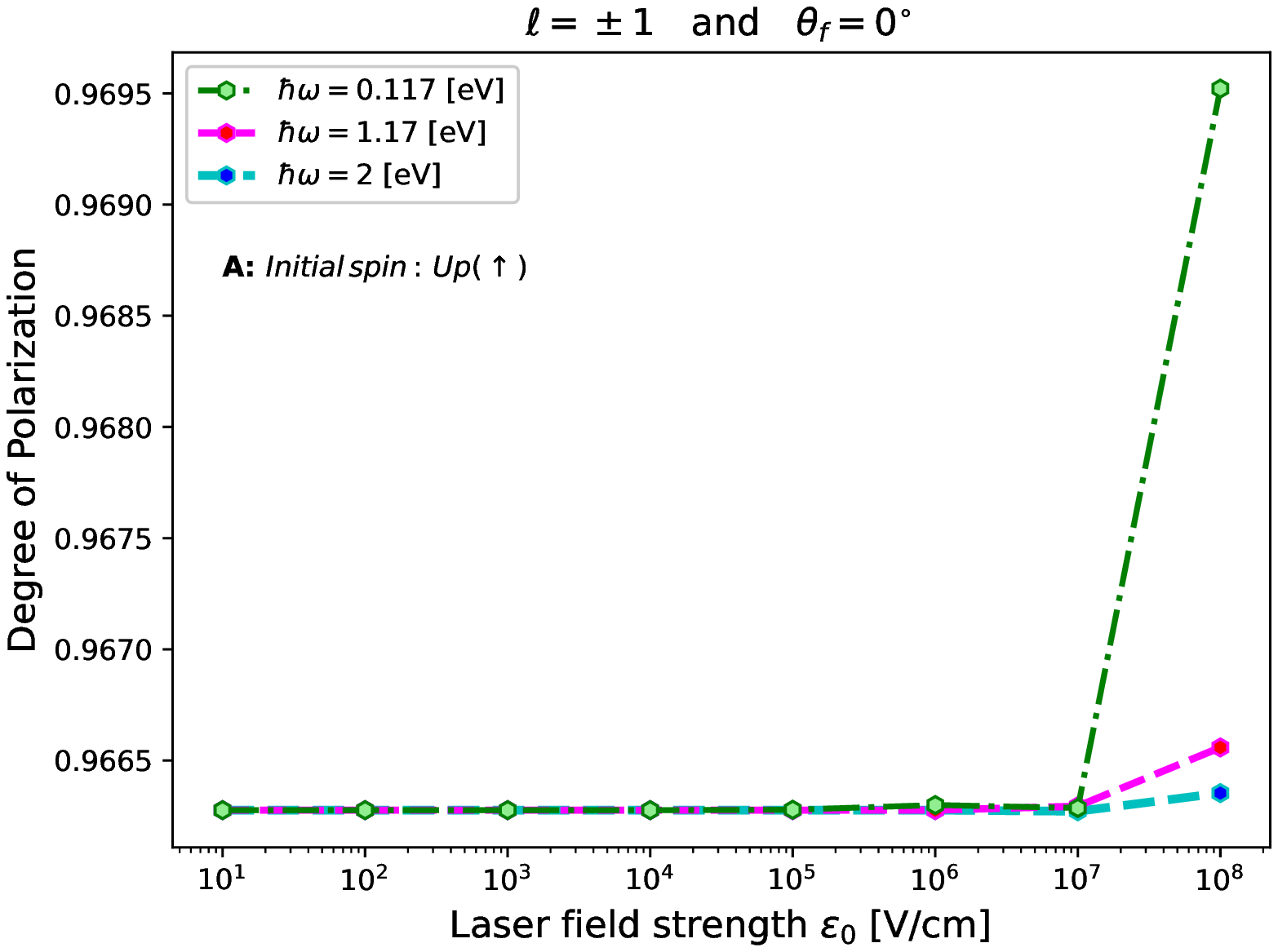}
  \end{minipage}
  \hspace*{0.25cm}
  \begin{minipage}[t]{0.42\textwidth}
  \centering
    \includegraphics[width=\textwidth]{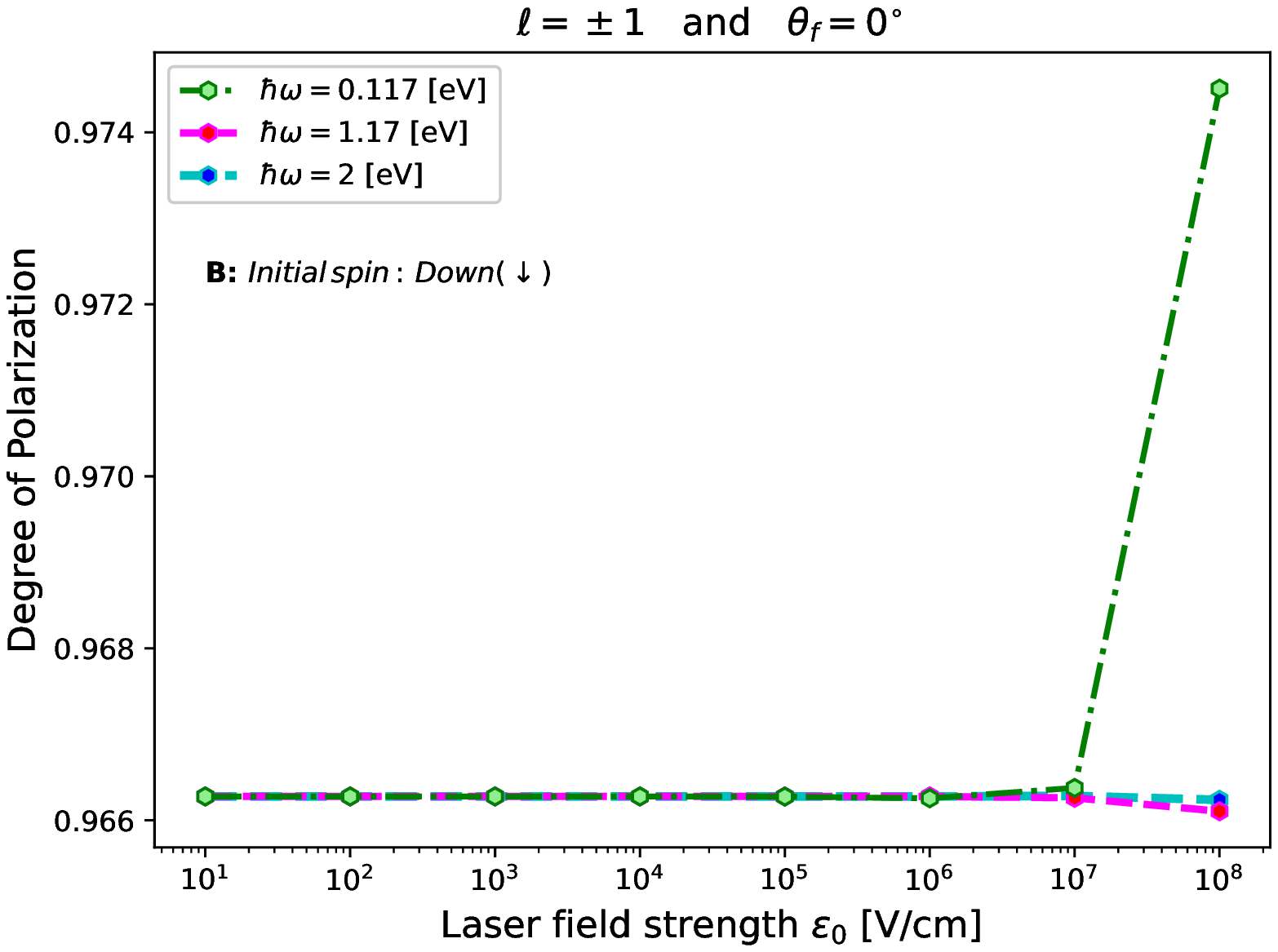}
  \end{minipage}
  \caption{Variation of the laser-assisted degree of polarization (DP) as a function of the field strength for different frequencies and for electrons initially in (A) spin Up and (B) spin Down.} \label{Figure:4}
\end{figure}
Finally, we present, in Fig.~\ref{Figure:4}, the effect of the laser field on the degree of polarization (DP) defined in Eq.~(\ref{DP}). Fig.~\ref{Figure:4}A depicts the variation of the degree of polarization as a function of the strength $ \varepsilon_{0} $ for electrons initially polarized in spin Up. We found that the degree of polarization slightly increased when we use a CO$_{2}$ laser. This increasing effect starts to appear from $ \varepsilon_{0}=10^{7}$ V/cm, where $\text{DP}=96,62\,\%$, up to $ \varepsilon_{0}=10^{8}$ V/cm where $\text{DP}=96,95\,\%$. For the other frequencies, the degree of polarization has increased less than before. In Fig.~\ref{Figure:4}B, where the electron is initially in spin Down, the degree of polarization increases this time from $96.63\,\%$ to $97.45\,\%$ respectively for $\varepsilon_{0}=10^{7}$ and $10^{8}$ V/cm in the case of a CO$_{2}$ pulsed laser, while it remains almost constant for other lasers. At the end, we conclude that the elastic e-p scattering process of initially well-defined electron spin assisted by laser field of frequency $ \hbar\omega=0.117$ eV is more probable for the scattering of non-flip electrons than for flip ones. Also, we conclude, as in \cite{Manaut_2005}, that the degree of polarization is weak depending on the number of photons exchanged $\ell$. 
\begin{figure}[H]
\centering
    \includegraphics[scale=0.51]{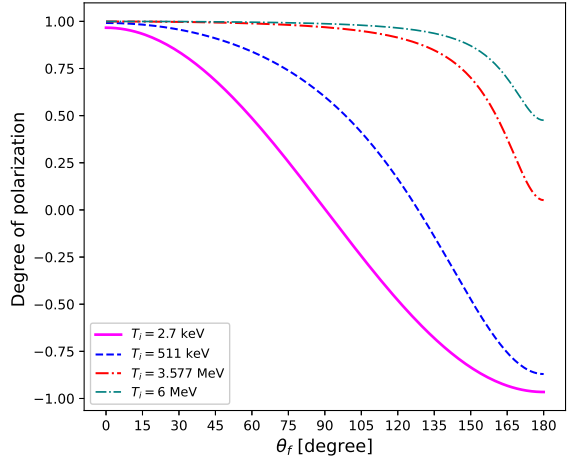}
\caption{Degree of polarization without laser as a function of scattering angle $\theta_f$ for different incident electron kinetic energies. The electron is considered to be initially in spin Up ($\uparrow$).} \label{figure51}
\end{figure}
To see the effect of the incident electron kinetic energy on the degree of polarization, we include in Fig.~(\ref{figure51}) its changes in terms of the final angle $\theta_f$ with respect to various kinetic energies. At low-energy scattering, the degree of polarization is strongly changed with the scattering angle $\theta_f$. On the other hand, at very high energy, when the scattering becomes relativistic, the degree of polarization becomes less dependent on the scattering angle $\theta_f$ and approaches the constant value $\text{DP}=1$, which implies the complete dominance of spin non-flip DCS. The physical meaning of this result is that at very high energy, the probability that the incident electron changes its spin is zero. These results are coherent with those obtained in previous articles \cite{idrissi2013,greiner} which follow the same methodology.
\section{Conclusion}\label{Sec3}
In this study, we have treated the process of laser-assisted electron-proton scattering considering the effect of electron spin polarization. By applying the concept of polarized electrons, the expressions of the DCS and
the degree of polarization are derived and their behavior in the presence of the laser field is studied. We found that the effect of high laser field strengths is important in each well defined polarization case (spin-flip and non-flip) for an exchanged number pf photons $\ell=\pm 1$. The comparison of the different spin polarizations, made by introducing the degree of polarization, shows that the laser field can modify both spin-flip and non-flip DCSs with a continuous dominance of the non-flip DCS. It is concluded that at high kinetic energy of the incident electron, the electron's probability for changing its spin is zero. A consistency check that is carried out successfully is that the sum of the two spin-flip and spin non-flip DCSs always gives the unpolarized DCS.  
This approach, developed to describe the electrons by Dirac-Volkov functions, is based on the fact that the involved proton is unpolarized and not dressed by the laser field. In order to provide an in-depth and comprehensive study, we are planning to also consider, in addition to the electron, the effect of proton dressing and its spin polarization in a future work.


\begin{thebibliography}{99}
\bibitem{Laser_technologies} European X-ray Free-Electron Laser (XFEL) \url{http://xfel.eu/}.

\bibitem{Laser_technologies_1} Linac Coherent Light Source II (LCLS II) \url{https://lcls.slac.stanford.edu/}.

\bibitem{Laser_technologies_2} Free-electron laser FLASH \url{https://flash.desy.de/}.

\bibitem{ioniz2s} M. Jakha, S. Mouslih, S. Taj, B. Manaut, M. El Idrissi, E. Hrour, E. Siher,
Relativistic electron-impact ionization of hydrogen atom from its metastable 2S-state in the symmetric/asymmetric coplanar geometries, Chin. J. Phys. \url{https://doi.org/10.1016/j.cjph.2021.09.005}

\bibitem{taj2019} S. Taj, B. Manaut, E. Hrour, M. El Idrissi, Laser-assisted positron-impact ionization of Hydrogen atoms, Acta Phys. Pol. A 136 (2019) 78.

\bibitem{idrissi2013}
M. El Idrissi, S. Taj, B. Manaut, Y. Attaourti, L. Oufni, Spin effects in laser-assisted semirelativistic excitation of atomic hydrogen by electronic impact, J. At. Mol. Sci. 4 (2013) 95.

\bibitem{taj2011} S. Taj, B. Manaut, M. El Idrissi, L. Oufni, Laser-assisted semi-relativistic excitation of atomic Hydrogen by electronic impact, Chin. J. Phys. 49 (2011) 1164.

\bibitem{manaut2009} B. Manaut, Y. Attaourti, S. Taj, S. Elhandi, Mott scattering of polarized electrons in a circularly polarized laser field, Phys. Scr. 80 (2009) 025304.

\bibitem{Manaut_2005} B. Manaut, S. Taj, Y. Attaourti, Mott scattering of polarized electrons in a strong laser field, Phys. Rev. A 71 (2005) 043401.

\bibitem{Attaourti} Y. Attaourti, B. Manaut, A. Makhoute, Relativistic electronic dressing in laser-assisted electron-hydrogen elastic collisions, Phys. Rev. A 69 (2004) 063407.

\bibitem{Dahiri_2021} I. Dahiri, M. Jakha, S. Mouslih, B. Manaut, S. Taj, Y. Attaourti, Elastic electron-proton scattering in the presence of a circularly polarized laser field, Laser Phys. Lett. 18 (2021) 096001.

\bibitem{Mekaoui2021} Y. Mekaoui, M. Jakha, S. Mouslih, B. Manaut, R. Benbrik, S. Taj,  Relativistic elastic scattering of an electron by a muon in the field of
a circularly polarized electromagnetic wave (Available at: \url{https://arxiv.org/abs/2110.06695})

\bibitem{Narozhny_2015} N.B. Narozhny, A.M. Fedotov, Extreme light physics, Contemp. Phys. 56 (2015) 249.

\bibitem{DiPiazza_2012} A. Di Piazza, C. Muller, K.Z. Hatsagortsyan, C.H. Keitel, Extremely high-intensity laser interactions with fundamental quantum systems, Rev. Mod. Phys. 84 (2012) 1177.

\bibitem{Ehlotzky_2009} F. Ehlotzky, K. Krajewska, J.Z. Kami\'{n}ski, Fundamental processes of quantum electrodynamics in laser fields of relativistic power, Rep. Prog. Phys. 72 (2009) 046401.

\bibitem{ElAsri2021} S. El Asri, S. Mouslih, M. Jakha, B. Manaut, Y. Attaourti, S. Taj, R. Benbrik, Elastic scattering of a muon neutrino by an electron in the presence of a circularly polarized laser field, Phys. Rev. D 104 (2021) 113001.

\bibitem{oualichin} M. Ouali, M. Ouhammou, Y. Mekaoui, S. Taj, B. Manaut, $Z$-boson production via the weak process $e^+ e^- \rightarrow \mu^+ \mu^-$ in the presence of a circularly polarized laser field, Chin. J. Phys. \url{https://doi.org/10.1016/j.cjph.2021.10.007}

\bibitem{pidecay} S. Mouslih, M. Jakha, S. Taj, B. Manaut, E. Siher,
Laser-assisted pion decay, Phys. Rev. D 102 (2020) 073006.

\bibitem{wdecay} M. Jakha, S. Mouslih, S. Taj, Y. Attaourti, B. Manaut Influence of intense laser fields on measurable quantities in $W^-$-boson decay, Chin. J. Phys. \url{https://doi.org/10.1016/j.cjph.2021.09.011}

\bibitem{kdecay} M. Baouahi, M. Ouali, M. Jakha, S. Mouslih, Y. Attaourti, B. Manaut, S. Taj, R. Benbrik, Laser-assisted kaon decay and CPT symmetry violation, Laser Phys. Lett. \textbf{18} (2021) 106001.

\bibitem{zdecay} M. Jakha, S. Mouslih, S. Taj, B. Manaut, Laser effect on the final products of $Z$-boson decay, Laser Phys. Lett. 18 (2021) 016002.

\bibitem{Ouali2021} M. Ouali, M. Ouhammou, S. Taj, B. Manaut, R. Benbrik Laser-assisted charged Higgs pair production in Inert Higgs Doublet Model (IHDM), Phys. Lett. B 823 (2021) 136761.

\bibitem{ouhammouchin} M. Ouhammou, M. Ouali, S. Taj, B. Manaut, Laser-assisted neutral Higgs-boson pair production in Inert Higgs Doublet Model (IHDM), Chin. J. Phys. \url{https://doi.org/10.1016/j.cjph.2021.09.012}

\bibitem{Ouhammou2021} M. Ouhammou, M. Ouali, S. Taj, B. Manaut, Higgs-strahlung boson production in the presence of a circularly polarized laser field, Laser Phys. Lett. 18 (2021) 076002.

\bibitem{Mourou_2011} G.A. Mourou, G. Korn, W. Sandner, J.L. Collier, ELI Whitebook \url{https://eli-laser.eu/media/1019/eli-whitebook.pdf}.

\bibitem{XCELS} Institute of Applied Physics RAS, Exawatt Center for Extreme Light Studies (XCELS) (2009) \url{http://xcels.iapras.ru/img/XCELS-Project-english-version.pdf}.

\bibitem{schwartz} M.L. Schwartz, Physics with polarized electron beams, Tech. Rep. SLAC-PUB-4656, Stanford Linear Accelerator Center, Stanford University, Stanford, CA (1988) 94309.

\bibitem{clendenin} J.E. Clendenin, Spin-Polarized Electrons: Generation and Applications, Tech. Rep., Stanford Linear Accelerator Center, Menlo Park, CA (US) (1999). 

\bibitem{hodge} L.A. Hodge, F.B. Dunning, G.K. Walters, Intense source of spin-polarized electrons, Rev. Sci. Instrum. 50 (1979) 1.

\bibitem{pierce1} D.T. Pierce, R.J. Celotta, G.C. Wang, W.N. Unertl, A. Galejs, C.E. Kuyatt, S. Mielczarek, The GaAs spin polarized electron source, Rev. Sci. Instrum. 51 (1980) 478.

\bibitem{pierce2} D.T. Pierce, F. Meier, P. Zurcher, Negative electron affinity GaAs: A new source of spin‐polarized electrons, Appl. Phys. Lett. 26 (1975) 670.

\bibitem{kessler} J. Kessler, Polarized Electrons, ed G Ecker et al., 2nd Edition, Springer, 1985.

\bibitem{anselmino95} M. Anselmino, A. Efremov, E. Leader, The theory and phenomenology of polarized deep inelastic scattering, Phys. Rep. 261 (1995) 1.

\bibitem{steven2009} S.D. Bass, The proton spin puzzle : where are we today ?, Mod. Phys. Lett. A 24 (2009) 1087.

\bibitem{prescott78} C.Y. Prescott, et al., Parity non-conservation in inelastic electron scattering, Phys. Lett. B 77 (1978) 347.

\bibitem{labzowsky2001} L.N. Labzowsky, A.V. Nefiodov, G. Plunien, G. Soff, R. Marrus, D. Liesen, Parity-violation effect in heliumlike gadolinium and europium, Phys. Rev. A 63 (2001) 054105.

\bibitem{compton1} T.V. Shishkina, A.L. Bondarev, Study of polarized effects in Compton scattering, Proc. of the F$\&$ANS-2010 Conf. -School 80 (2010).

\bibitem{compton2} G.L. Kotkin, S.I. Polityko, V.G. Serbo, Polarization of final electrons in the Compton effect, Nucl. Instr. and Meth. in Phys. Res. A 405 (1998) 30.

\bibitem{compton3} F.W. Lipps, H.A. Tolhoek, Polarization phenomena of electrons and photons. I: General method and application to Compton scattering, Physica 20 (1954) 85.

\bibitem{gakh} G.I. Gakh, A. Dbeyssi, D. Marchand, E. Tomasi-Gustafsson, V.V. Bytev, Polarization effects in elastic proton-electron scattering, Phys. Rev. C 84 (2011) 015212. 

\bibitem{volkov} D.M. Volkov, On a class of solutions of the Dirac equation, Z. Phys. 94 (1935) 250.

\bibitem{greiner} W. Greiner, J. Reinhardt, Quantum Electrodynamics, 4th Edition, Springer, 2009.

\bibitem{Abramowitz} M. Abramowitz, I. Stegun, Handbook of Mathematical Functions, Dover, New York, 1970.

\bibitem{feyncalc} V. Shtabovenko, R. Mertig, F. Orellana, FeynCalc 9.3: New features and improvements, Comput. Phys. Commun. 256 (2020) 107478.

\bibitem{andrick} D. Andrick, L. Langhans, Measurement of free-free transitions in e--Ar scattering, J. Phys. B: At. Mol. Opt. Phys 9 (1976) L459.

\bibitem{Weingartshofer_1977} A. Weingartshofer, J.K. Holmes, G. Caudle, E.M. Clarke, H. Kruger, Direct observation of multiphoton processes in laser-induced free-free transitions, Phys. Rev. Lett. 39 (1977) 269.

\bibitem{kroll} N.M. Kroll, K.M. Watson, Charged-particle scattering in the presence of a strong electromagnetic wave, Phys. Rev. A 8 (1973) 804.

\end{thebibliography}
\end{document}